\documentclass[12pt,english]{iopart}
\usepackage{iopams}

\usepackage{color}
\usepackage[pdftex]{graphicx,hyperref}% Include figure files

\begin{document}

\title{Optical conductivity of ABA stacked graphene trilayer: mid-IR resonance due to band nesting}

\author{Zeinab Rashidian$^{1,2}$, Yuliy V. Bludov$^{1}$, Ricardo M. Ribeiro$^{1}$, N. M. R. Peres$^{1}$, Mikhail I. Vasilevskiy$^{1}$}

\address{$^{1}$Centro de F\'{i}sica and Departamento de F\'{i}sica, Universidade
do Minho, Campus de Gualtar, Braga 4710-057, Portugal}

\address{$^{2}$Department of Physics, Faculty of Science, Lorestan university(lu), Lorestan, Iran
}

\ead{mikhail@fisica.uminho.pt}

%\date{\today}

\begin{abstract}
The band structure and the optical conductivity of an ABA (Bernal-type) stacked graphene
 trilayer are calculated. It is shown that, under appropriate doping, a strong resonant peak
develops in the optical conductivity, located at
the frequency corresponding to approximately 1.4 times the interlayer hopping energy
and caused by the "nesting" of two nearly parabolic bands in the electronic spectrum.
The intensity of this resonant absorption can be controlled by adjusting the gate voltage.
The effect is robust with respect to increasing temperature.
\end{abstract}

\pacs{81.05.ue, 72.80.Vp, 78.67.W}
%\keywords{graphene, Bernal stacking, optical absorption}

\maketitle

\section{Introduction}
Since the isolation of monolayer graphene almost a decade ago~\cite{novoselov},
there has been a high interest in the low energy  transport  and optical properties
of not only the monolayer~\cite{neto,Grigorenko} but also a few layer
 graphene systems.~\cite{nilson,kuzmenko,zhang,duppen,koshino1}
These properties are determined by the electronic band
structure near the $\mathbf K$ point. The undoped monolayer  graphene (MLG)
 is characterized by the universal optical conductivity, $\sigma _0=e^2/(4\hbar)$.
This implies that the transmittance
 depends solely on the fine structure constant and originates the quantized
visible opacity of suspended monolayer graphene~\cite{nair,stauber}.
As far as doped graphene is concerned, there are several effects that arise
in the optical properties, related to the restrictions introduced on the
interband transitions by the state filling and also to the onset of intraband
transitions~\cite{Falkovsky}. The latter correspond to plasmons and give rise to the interesting
 and promising field of graphene plasmonics~\cite{Grigorenko,PRIMER}.

Graphene multilayers offer a new ingredient to the interesting physics and
potential applications. The relatively weak interlayer coupling, on the one hand,
implies that they should inherit some properties of the parent material~\cite{nilson},
on the other hand, it introduces a new energy scale, of the order of few
 tenths of the electron-volt, that should yield some new properties.
It has been shown~\cite{macdonald2} that there is  also a universal optical conductivity
in a undoped $N$-layer graphene, equal to $\sigma_{0N}=N\sigma_{0}$, that is
reached in undoped graphene in the low frequency limit.
At the same time, the optical response of doped bilayer graphene reveals intense strongly doping-dependent features in the mid-infrared (around 0.4~eV)~\cite{kuzmenko,zhang}. The origin of these experimentally observed features has been considered theoretically in these works and also, in more detail, in Ref. ~\cite{carbott}, where the band structure and the optical conductivity of bilayer graphene were calculated. Some novel plasmonic effects in Bernal-stacked bilayer graphene were predicted in the recent work~\cite{Low-archive}.
As the number of layers increases beyond $N=2$, the band structure and the optical conductivity become dependent not only on $N$ but also upon the stacking arrangement.
There are three distinct planar projections of the honeycomb lattice (usually denoted A, B and C) and, consequently, $2^{N-2}$ distinct $N-$layer sequences~\cite{macdonald2}. In particular, the
stacking of three layers in a graphene trilayer can be either ABA (also called Bernal-type) and ABC (rhombohedral)~\cite{koshino}.
These two different stacking arrangements lead to strikingly different electronic band strucures~\cite{Menezes2014}.
For instance, it was found that undoped graphene ABC trilayer
shows many-body correlations with an energy gap, while the Bernal-type
 stacking (taking place in graphite) does not lead to a gap~\cite{scherer}.
Non-Bernal stacked multiple graphene layers have attracted a considerable
attention related to the prospect of further enhancement of capabilities
of graphene-based optoelectronic devices, in particular, THz and IR photodiodes~\cite{Ryzhii}. The optical conductivity of ABC trilayers has been considered in a number of works~\cite{macdonald1,Morimoto2012,Xiao2013}, in particular, the effect of doping has been analysed~\cite{Xiao2013}. Even though trilayer graphene contains, on average, regions of ABC and ABA stacking in an 15:85 ratio~\cite{Henriksen2012}, apparently the optoelectronic properties of the latter attracted less attention and we are aware of only one work~\cite{Ubrig2012} devoted to this topic. This is notwithstanding the possibility of using far-infrared (FIR) spectroscopy, along with the common Raman scattering technique~\cite{Raman} in order to distinguish different trilayer graphene species~\cite{Henriksen2012,Ubrig2012}.

Therefore, the main purpose of this article is to analyse the spectral
characteristics of the optical conductivity of intrinsic and doped ABA stacked trilayers.
 The mirror-symmetric Bernal stacking is the most common in graphene multilayers that
can be exfoliated from natural graphite since it shares its
 crystalline structure~\cite{duppen}. Intuitively, one can
 expect that the electrons in ABA trilayers can have a monolayer- or bilayer-like
 character~\cite{Orlita-Potemski}. We shall present the analytical dispersion relation
for the electrons near the Dirac point, from which it follows that the band structure of the
ABA trilayer indeed looks like a superposition of those characteristic of a monolayer and a bilayer,
although the latter corrseponds to an effective interlayer hopping constant $\sqrt {2}$
times larger than the true one.
Based on this band structure, we calculate both the interband and intraband (Drude)
contributions to the optical conductivity. The most interesting result is that the
optical absorption of a doped ABA trilayer is dominated by a narrow
resonant peak at the frequency corresponding to this effective interlayer
hopping energy (approximately 0.56~eV). It is caused by the fact that the
dispersion curves corresponding to two bands are nearly parallel for a
considerable range of wavevectors near the Dirac point. This effect sometimes
is called "band nesting"~\cite{Carvalho} in order to distinguish from van
Hove singularities in the single-particle density of states. We will show that
the intensity of this resonant absorption is approximately proportional to the
Fermi energy and, therefore, can be controlled by adjusting the gate voltage
applied to the graphene layer.

\section{Theoretical Background}
\subsection{Band structure}

\begin{figure}[b]
\begin{center}
\includegraphics[width=10cm]{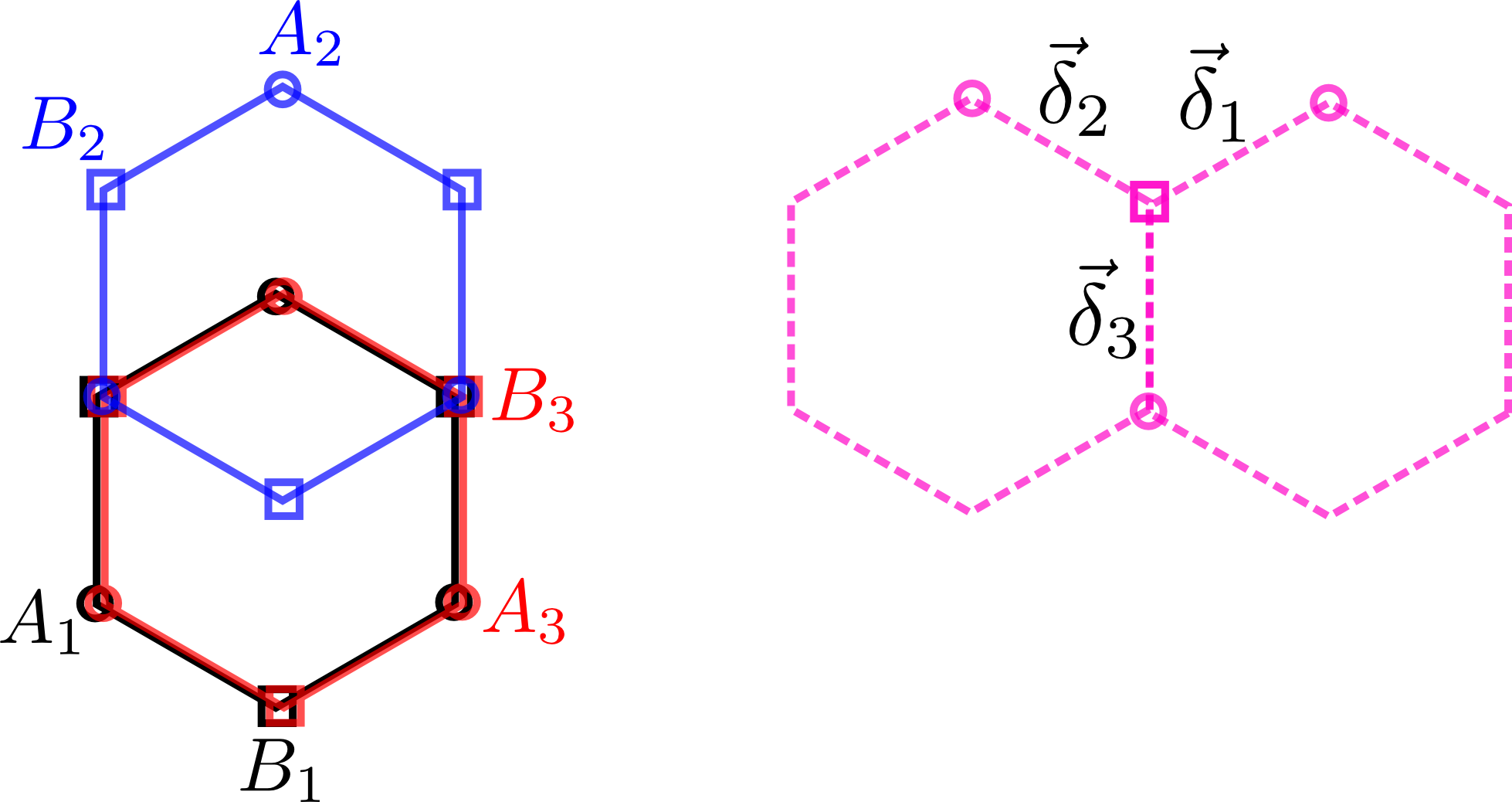}
\end{center}
\caption{Lattice structure of trilayer graphene with ABA stacking (top view).
{It can be seen as a hexagonal Bravais lattice with a six-atom basis (one A and one B atom for each layer). The right panel shows three nearest neighbours (A atoms) of a B atom. }
}
\label{fig:f1}
\end{figure}

The tight-binding Hamiltonian for non-interacting electrons in the ABA stacked trilayer involves three A-type and three B-type sites and includes the essential in-plane ($t_0\approx 2.7\,$eV) and  interlayer hoppings ($t_1\approx 0.4\,$eV) as shown in Fig.~\ref{fig:f1}. {These two parameters, connecting atoms that are right on top of each other in adjacent layers, are sufficient to describe the main features of the band structure, such as the type of dispersion of the energy bands and their separation, as confirmed by recent DFT calculations~\cite{Menezes2014}. Keeping only $t_0$ and $t_1$ hoppings permits to obtain simple formulae for band gaps, effective masses and the Fermi velocity.}
Then the Hamiltonian is given by~\cite{scherer}:
%\begin{widetext}
\begin{eqnarray}
\nonumber
\hat{H}=-t_{0}\sum_{n,\delta_{j}}\vert A_{1},\vec R_{n}+\vec \delta_{j}\rangle \langle B_{1}, \vec R_{n} \vert \\
\nonumber
-t_{0}\sum_{n,\vec \delta_{j}} \vert A_{3},\vec R_{n}+\vec \delta_{j}\rangle \langle B_{3},\vec R_{n}\vert
\nonumber
-t_{0}\sum_{n,\vec \delta_{j}}\vert A_{2},\vec R_{n}\rangle \langle B_{2},\vec R_{n}-\vec \delta_{j}\vert \\
%\nonumber
+t_1\sum_{n}\vert A_{2},\vec R_{n}\rangle \langle B_{1},\vec R_{n}\vert
+t_1\sum_{n}\vert A_{2},\vec R_{n}\rangle \langle B_{3},\vec R_{n}\vert \:+ \: \mathbf {H. c.}
\label {Ham}
\end{eqnarray}
%\end{widetext}
Here $n=(n_1,n_2)$ is the composite index, which determines the atomic positions in the lattice, $\vec R_{n}=n_1\vec g_1+n_2\vec g_2$, with $\vec g_1=(\frac{\sqrt{3}}{2},\frac{3}{2})a_0$, $\vec g_2=(-\frac{\sqrt{3}}{2},\frac{3}{2})a_0$ being the lattice vectors. The three vectors that connect the $B$ atom to its three nearest neighbors are
$\vec\delta_{1}=(\frac{\sqrt{3}}{2},-\frac{1}{2})a_0$,
$\vec\delta_{2}=(-\frac{\sqrt3}{2},-\frac{1}{2})a_0$,
and $\vec\delta_{3}=(0,1)a_0$, where $a_0$ is the C-C interatomic distance.
The positions of the $A$ atoms relative to the $B$
atoms in each of the three layers are shown in Fig.~\ref{fig:f1}.

\begin{figure}[!ht]
\begin{center}
\includegraphics[width=10 cm]{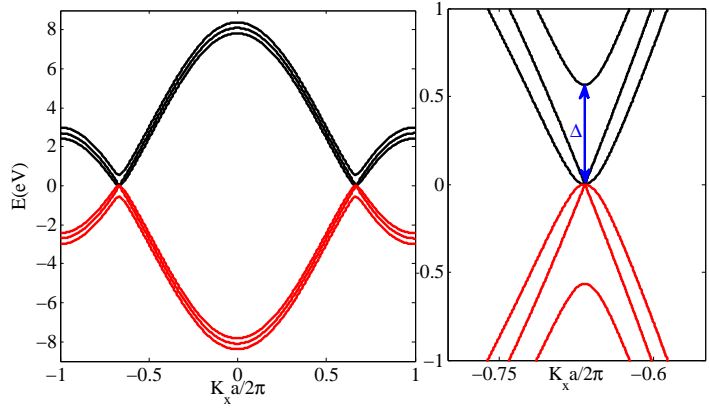}
\end{center}
\caption{Band structure of ABA stacked trilayer graphene {throught the Brillouin zone (left) and near the $\mathbf K_+$ Dirac point (right), where $\Delta=\sqrt2 t_1 $. The $k_x$ axis is scaled by $2\pi /a$, $a=a_0\sqrt 3$ is the lattice constant. After intersecting the $\mathbf K_{\pm}$ points the dispersion curves continue along the $\mathbf {KM}$ direction. }
}
\label{fig:f2}
\end{figure}

The energy spectrum of the Hamiltonian (\ref {TB-Hamiltonian}) is composed of six energy bands given by (see Appendix A for details):
\begin{eqnarray}
E_{\pm1}\left(\vec k\right)=\pm\sqrt{t_1^2+\vert \phi\left(\vec k\right)\vert^2- t_1\sqrt{t_1^2+2\vert \phi\left(\vec k\right)\vert^2}}\:;
\label{eq:e1}
\end{eqnarray}
\begin{eqnarray}
E_{\pm2}\left(\vec k\right)=\pm\sqrt{\vert \phi\left(\vec k\right)\vert^2}\:;
\label{eq:e2}
\end{eqnarray}
\begin{eqnarray}
E_{\pm3}\left(\vec k\right)=\pm\sqrt{t_1^2+\vert \phi\left(\vec k\right)\vert^2+ t_1\sqrt{t_1^2+2\vert \phi\left(\vec k\right)\vert^2}}\:,
\label{eq:e3}
\end{eqnarray}
where
\begin{equation}
\phi(\vec{k})=-t_{0}\left [\exp\left(ik_{y}a_{0}\right)+2\exp\left(\frac{-ik_{y}a_{0}}{2}\right)\cos
\left(\frac{k_{x}\sqrt3}{2}a_{0}\right)\right]\:.
\label {eq:phi}
\end{equation}
The band structure is depicted in Fig.~\ref{fig:f2}. Note that the gap between the bands $\pm3$ and the Dirac point is ${2\Delta=2\sqrt2 }t_1$, while its counterpart in graphene bilayer is just equal to $2t_1$.~\cite{carbott}

In the vicinity of $\mathbf K_{\pm }$ points [Dirac points, $\vec k=\pm (4\pi/(3\sqrt{3}a_0),0)$] the band
energies can be approximated in the following way:
\begin{eqnarray}
\label{eq:parab}
%\begin{aligned}
E_{\pm1}=\pm\frac{\hbar^2K^2}{2m_{1}}\:;
\ E_{\pm2}=\pm\hbar \acute{v_{f}}K\:;
\ E_{\pm3}=\pm\left(\frac{\hbar^2K^2}{2m_{3}}+\Delta\right)\:,
%\end{aligned}
\end{eqnarray}
where $K=\sqrt{k_y^2+k_y^2}$.
Note that the effective masses at the bottom of the bands 1 and 3 are equal within the present model, $m_1=m_3=t_1/(\sqrt2 v_f^2)$, and the Fermi velocity of the Dirac-type band coincides with that of monolayer graphene, $\acute{v_{f}}=3t_{0}a_{0}/(2\hbar)=v_f$.

\subsection{Optical conductivity: interband part}

The finite frequency (optical) conductivity is calculated
through a standard procedure using the Kubo formula~\cite{cubo},
\begin{eqnarray}
\sigma (\omega)=\frac{-2ie^2}{\omega S}\sum_{\vec k,l^\prime\neq l}\left|\left\langle \vec k,l\right|\hat{V}_x\left|\vec k,l^\prime\right\rangle\right|^2
\frac{n_{f}\left[E_{l}(\vec k)\right]-n_{f}\left[E_{l^\prime}(\vec k)\right]}{\hbar\omega-E_{l}(\vec k)+E_{l^\prime}(\vec k)+i\Gamma}
\:,
\label {Kubo}
\end{eqnarray}
where $e$ is the electron charge, $S=\frac{3\sqrt{3}a_{0}^{2}}{2}$ is the unit cell area, $n_{f}$ is the
Fermi function, $\left|\vec k,l\right\rangle$ is the eignestate corresponding to energy $E_{l}(\vec k)$ (${l,l^\prime}$ are the band indices), $\Gamma$ is a damping parameter and $\hat{V}_x$ denotes the $x$-component of in-plane velocity operator, defined by
%\begin{widetext}
\begin{eqnarray}
\nonumber
\hat{\vec{V}}=\frac{it_{0}}{\hbar}\sum_{\delta_{j}}\vec\delta_{j}\left[\left|B_{1},\vec R_{n}\right\rangle\left\langle A_{1},\vec R_{n}+\vec \delta_{j}\right|-\left|A_{1},\vec R_{n}+\vec \delta_{j}\right\rangle\left\langle B_{1},\vec R_{n}\right|\right.\\
\nonumber
+\left|B_{2},\vec R_{n}-\vec\delta_{j}\right\rangle\left\langle A_{2},\vec R_{n}\right|-
\left|A_{2},\vec R_{n}\right\rangle\left\langle B_{2},\vec R_{n}-\vec \delta_{j}\right|\\
\left.+\left|B_{3},\vec R_{n}\right\rangle\left\langle A_{3},\vec R_{n}+\vec \delta_{j}\right|-\left|A_{3},\vec R_{n}+\vec \delta_{j}\right\rangle\left\langle B_{3},\vec R_{n}\right|\right]\:.
\end{eqnarray}
%\end{widetext}
The summation over $\vec k$ in Eq. (\ref {Kubo}) involves wavevectors in the first Brilluion zone, {which is the usual hexagon with side $\frac{4\pi}{(3\sqrt 3 a_{0})}$. In practice, this summation is replaced by a 2D integration over the triangle formed by the points $\mathbf \Gamma$ $(0;\:0)$, $\mathbf K$ $(\frac{4\pi}{(3\sqrt 3 a_{0})};\:0)$, and $\mathbf M$ $(\frac{\pi}{(3 a_{0})};\:\frac{\pi}{(\sqrt 3 a_{0})})$, and the result is multiplied by twelve. }

\subsection{Optical conductivity: Drude part}

For doped graphene ($\mu \neq 0$) it is necessary to include in the optical conductivity also the term related to intraband transitions, often referred to as Drude term.
Although it can be calculated through the Kubo formula, it is easier to derive this term using the Boltzmann transport
equation~\cite {PRIMER,Peres-RMP}. Thus, the electric current is written as
\begin{equation}
\vec{J}=\frac{4e}{(2\pi)^{2}}\sum_{l=1}^3\int d\vec{k}\,\,\delta n_{\vec{k}}^{(l)}\,\vec{v}_l(\vec{k})\,,
\label{eq:semi-classical_current}
\end{equation}
where $\delta n_{\vec{k}}^{(l)}$ is the deviation of the carriers distribution from the equilibrium Fermi--Dirac function, $n_{f}[E_l(\vec k)]$. The former is readily obtained from the Boltzmann equation,
\begin{equation}
\delta n_{\vec{k}}^{(l)}=\frac{e\vec{\mathcal E}\cdot \vec{v}_l(\vec{k})}{{\tau(\vec{k})}^{-1}-i\omega}\left (-\frac {\partial n_{f}[E_l(\vec k)]}{\partial E_l(\vec k)} \right )\:,
\label{eq:n_k}
\end{equation}
where $\vec{v}_l(\vec{k})$ is the group velocity of the charge carriers in the $l$-th band and $\tau(\vec{k})$ denotes the carrier relaxation time. The factor of 4 in Eq. (\ref{eq:semi-classical_current}) is due to the spin and valley degeneracy.

Substitution of Eq. (\ref{eq:n_k}) into Eq. (\ref{eq:semi-classical_current}) yields the Drude conductivity and for zero temperature we have:
\begin{eqnarray}
%\nonumber -\frac {e^2}{\pi \hbar^2}\sum _{i=1-3} \int\frac{\partial{f_0}}{\partial{\epsilon}} \frac{\tau_p(\epsilon)}{1-i\omega\tau_p(\epsilon)}v^2pd_p\\
\sigma _D =\frac{4\sigma_0}{\pi}\left [\frac{3\mu}{\hbar(\gamma-i\omega)}
+\frac{2(\mu-\Delta)}{\hbar(\Gamma-i\omega)}\theta(\mu-\Delta)\right ]\:,
\label {Drude}
\end{eqnarray}
where the damping parameters $\gamma$ and $\Gamma$ are defined as the inverse of the corresponding relaxation time at the Fermi level (we make no distinction between bands 1 and 2). This Drude term has to be added to the optical conductivity (\ref {Kubo}). Note that the first term is precisely three times the Drude conductivity of monolayer graphene.

\section{Results and Discussion}

\begin{figure}[!ht]
\begin{center}
\includegraphics[width=10cm]{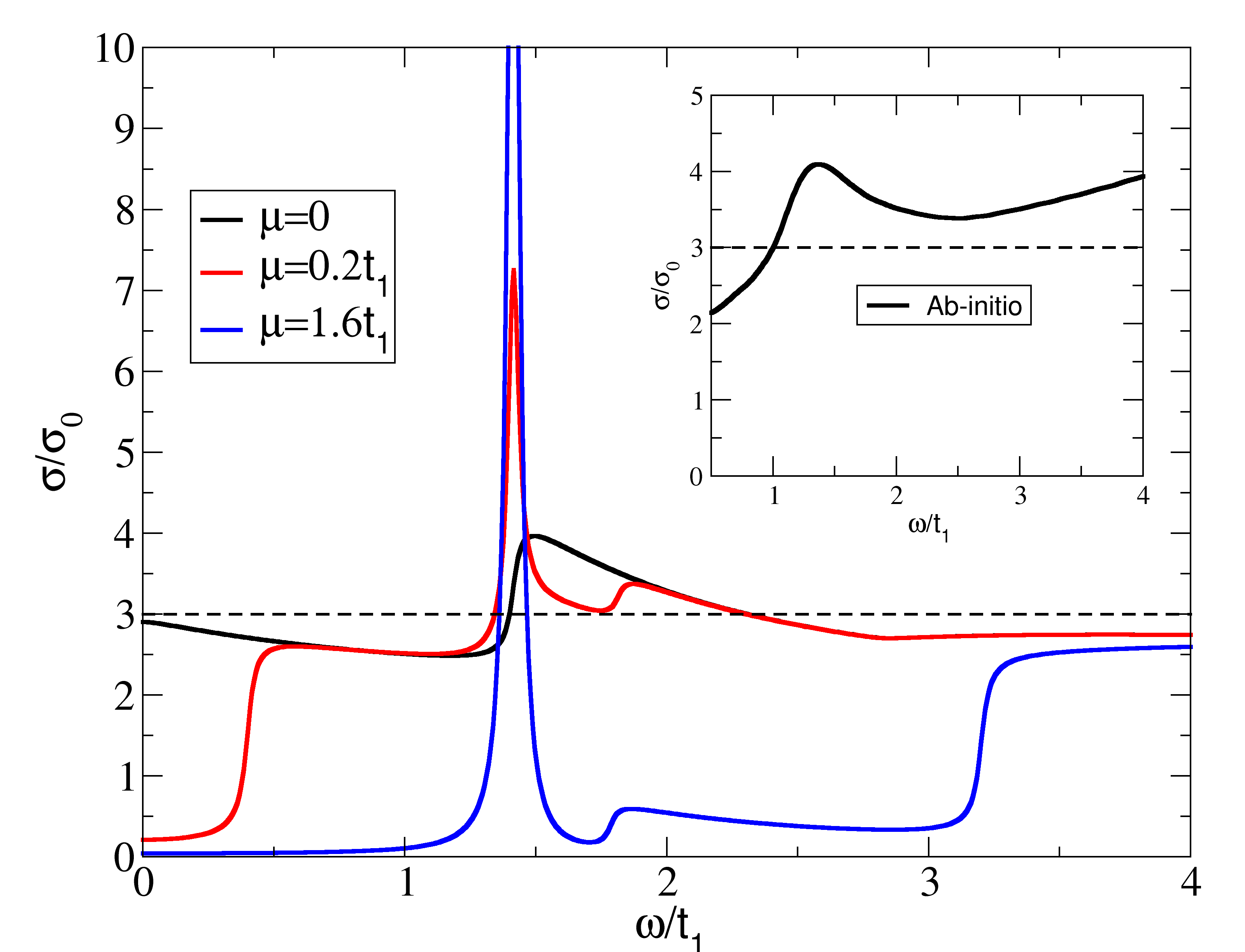}
\end{center}
\caption{Real part of optical conductivity of ABA
stacked trilayer graphene for $T=1K$.
Thin dashed line indicates the value $3\sigma_0$.
The inset shows the real part of the optical conductivity of an intrinsic ABA stacked trilayer graphene
computed using {\it ab-initio} methods~\cite {ab-initio}.
The peak is located where the tight-binding calculation predicts it.
}
\label{fig:f3}
\end{figure}

\begin{figure}[!ht]
\begin{center}
\includegraphics[width=10 cm]{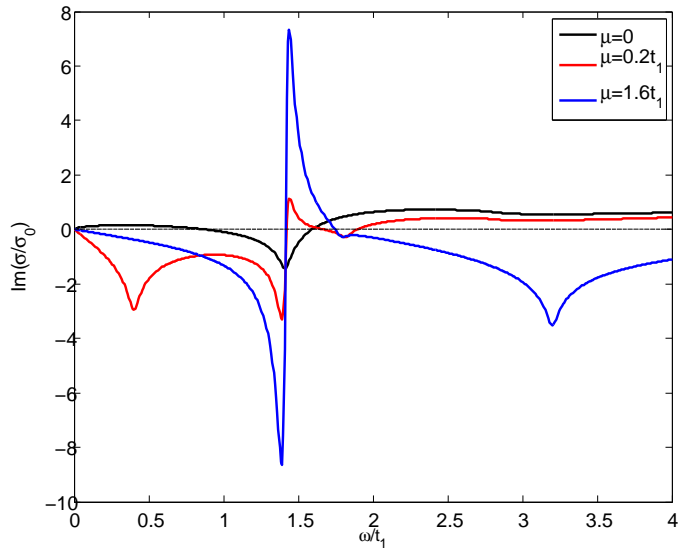}
\end{center}
\caption{Imaginary part of optical conductivity of ABA stacked trilayer graphene for $T=1K$.}% The inset shows the optical conductivity with just taking into account the transition from band 1 to 3.}
\label{fig:f4}
\end{figure}

We shall now concentrate on the frequency dependence of the real part of the derived optical conductivity, $\sigma ^\prime(\omega)$, that determines the absorption,
for different values of the chemical potential, $\mu$.
The spectra of real, $\sigma ^\prime(\omega)$, {and imaginary, $\sigma ^{\prime\prime}(\omega)$, parts of the conductivity} for three different values of $\mu$ (conveniently expressed in units of $t_1$) are presented in Figs. \ref{fig:f3} and \ref{fig:f4}, {respectively}. Note that the first value corresponds to undoped graphene ($\mu=0$), the second one is $\mu=0.2t_1<\Delta$ and the last $\mu>\Delta$.
For intrinsic trilayer graphene, either ABA or ABC stacked, $\sigma ^\prime(\omega)$ tends to $3\sigma _0$ (\cite {Xiao2013,Ubrig2012}).
We clearly see the effects of doping present as the Fermi step  at $\hbar\omega=2\mu$ (for $\mu>0$) , known in semiconductors as Burstein--Moss effect \cite {Grundmann} and familiar in monolayer graphene~\cite {Peres-RMP}. The feature characteristic of trilayer graphene, located at $\hbar\omega\geq\Delta$, splits into two for $\mu>0$ (compare red and {blue} curves to the black one in  Fig.~\ref{fig:f3} ). But the most impressive effect of the doping is the onset of the resonant peak at $\hbar\omega =\Delta$, whose intensity grows strongly with $\mu$.

In order to understand these spectral changes caused by doping, we analyzed all possible optical transitions listed in Table 1 (note that none of the velocity matrix elements vanishes, they are all allowed!).
As it can be seen from Fig.~\ref{fig:f2}, there are two Dirac-type and four approximately parabolic bands that arise from six atoms in the unit cell of the ABA trilayer. Let us consider the allowed transitions that correspond to the features of the optical conductivity. For the undoped graphene ($\mu=0$), we have nine possible transitions, with energy conservation restrictions imposed on some of them (see Table 1).
There are four transitions allowed without any restriction for all frequencies, namely, $l^\prime=-1$ to $l=1$ (denoted as $-1 \rightarrow 1$, $-1 \rightarrow 2$,
$-2 \rightarrow 1$ and $-2 \rightarrow 2$.
The onset of transitons involving bands $\pm3$ is at $\hbar\omega>\sqrt2t_1$ and  $\hbar\omega>2\sqrt2t_1$. That is why there are four possible transitions including $-1 \rightarrow 3$ , $-2 \rightarrow 3$, $-3 \rightarrow 1$ and $-3 \rightarrow 2$, with a threshold frequency $\hbar\omega=\sqrt2{t_1}$. Still another possible transition is from $-3$ to $3$, with a threshold at
$\hbar\omega=2\sqrt2t_1$, although it is less pronounced as clearly seen in Fig.~\ref{fig:f3} (bold black curve). With doping the system so as $\mu=0.2{t_1}$, the behavior of the optical conductivity is altered, caused by the changes in the allowed transitions with applying more state--filling restrictions as well as the onset of new transitions.

\begin{figure}[!ht]
\begin{center}
\includegraphics[width=10 cm]{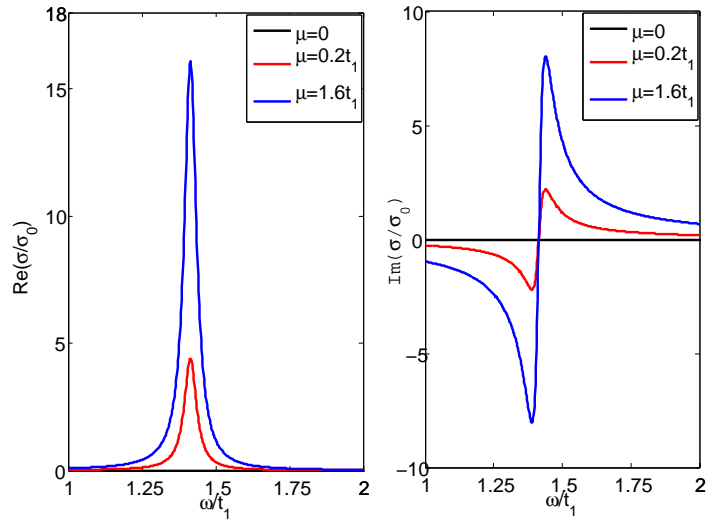}
\end{center}
\caption{Real and imaginary part of the optical conductivity including only the $1 \rightarrow 3$ transition ($T=1K$).}% The inset shows the optical conductivity with just taking into account the transition from band 1 to 3.}
\label{fig:f12}
\end{figure}

\begin{figure}[!ht]
\begin{center}
\includegraphics[width=10 cm]{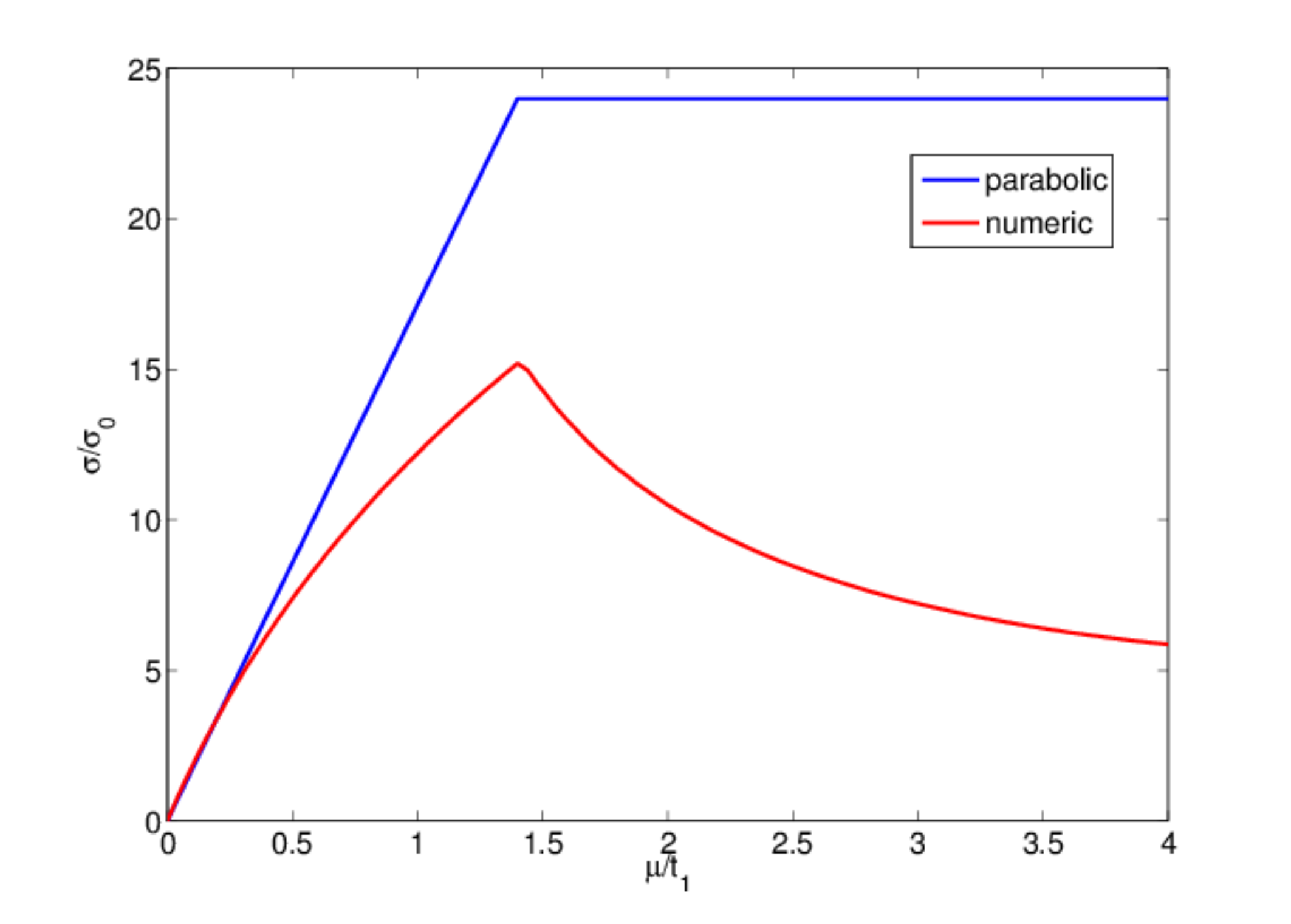}
\end{center}
\caption{Relative intensity of the $1 \rightarrow 3$ transition {\it versus} Fermi energy, calculated analytically within parabolic approximation and numerically.}
\label{fig:f5}
\end{figure}

The general expression for the threshold energy for the $l^\prime \rightarrow l$ transition {(for the case $l^\prime<0$, $l>0$)} is given by:
\begin{eqnarray}
\nonumber
\Delta_{l^\prime \rightarrow l}=E_{l}\left(K_{f}^{(l)}\right)-E_{l^\prime}\left(K_{f}^{(l)}\right),
\end{eqnarray}
where $K_{f}^{(l)}$ is the Fermi wavevector in the $l$-th band (the root of equation $E_{l}\left(K_{f}^{(l)}\right)=\mu$, or $K_{f}^{(l)}=0$, if the equation does not have a real root).
{At the same time, for the case where $l^\prime,\:l>0$ the expression for the threshold energy is given by
\begin{eqnarray}
\Delta_{l^\prime \rightarrow l}={\rm min}\left[E_{l}\left(K_{f}^{(l)}\right)-E_{l^\prime}\left(K_{f}^{(l)}\right),\:E_{l}\left(K_{f}^{(l^\prime)}\right)-E_{l^\prime}\left(K_{f}^{(l^\prime)}\right)\right].
\nonumber
\end{eqnarray}
There is also an upper cut-off frequency in this case, given by
\begin{eqnarray}
\Omega_{l^\prime \rightarrow l}={\rm max}\left[E_{l}\left(K_{f}^{(l)}\right)-E_{l^\prime}\left(K_{f}^{(l)}\right),\:E_{l}\left(K_{f}^{(l^\prime)}\right)-E_{l^\prime}\left(K_{f}^{(l^\prime)}\right)\right],
\nonumber
\end{eqnarray}
so that the allowed transition frequencies lie inside the domain $\Delta_{l^\prime \rightarrow l}\le\hbar\omega\le \Omega_{l^\prime \rightarrow l}$.}
It is possible to obtain simple expressions for the threshold energies using the approximation (\ref{eq:parab}). Thus, we obtain the following values:
\begin{eqnarray}
\Delta_{-1 \rightarrow 1}=2\mu,\label{eq:trans_m1_to_1}\ \ \Delta_{-1 \rightarrow 2}=\mu+\frac{\mu^2}{\sqrt2t_1}, \label{eq:trans_m1_to_2}\\
\Delta_{-1 \rightarrow 3}=2\left(\mu-\sqrt2t_1\right)\theta(\mu-\sqrt2t_1)+\sqrt2t_1, \label{eq:trans_m1_to_3}
\end{eqnarray}

\begin{eqnarray}
\Delta_{-2 \rightarrow 1}=\mu+\sqrt{\sqrt2t_1\mu},\label{eq:trans_m2_to_1} \ \ \Delta_{-2 \rightarrow 2}=2\mu,\\
\label{eq:trans_m2_to_2}
\Delta_{-2 \rightarrow 3}=\sqrt2t_1+\theta(\mu-\sqrt2t_1)\times
\left(\mu-\sqrt2t_1+\sqrt{\sqrt2t_1\mu-2t_1^2}\right),\label{eq:trans_m2_to_3}
\end{eqnarray}

\begin{eqnarray}
\Delta_{-3 \rightarrow 1}=2\mu+\sqrt2{t_1},\label{eq:trans_m3_to_1}\ \ \Delta_{-3 \rightarrow 2}=\mu+\sqrt2{t_1}+\frac{\mu^2}{\sqrt2{t_1}},\label{eq:trans_m3_to_2}\\
\Delta_{-3 \rightarrow 3}=2\left(\mu-\sqrt2{t_1}\right)\theta(\mu-\sqrt2t_1)+2\sqrt2{t_1}.\label{eq:trans_m3_to_3}
\end{eqnarray}
Several transitions can occur only for nonzero $\mu$:
\begin{eqnarray}
\Delta_{1 \rightarrow 2}=\mu-\frac{\mu^2}{\sqrt2t_1},\label{eq:trans_1_to_2}\ \ \Delta_{1 \rightarrow 3}=\sqrt2,\label{eq:trans_1_to_3}\\
\Delta_{2 \rightarrow 3}=\sqrt2t_1+\frac{\mu^2}{\sqrt2{t_1}}-\mu.\label{eq:trans_2_to_3}
\end{eqnarray}

\begin{table}[ht]
\caption{Optical transitions and threshold frequencies in units of $t_1$.} % title of Table
\centering % used for centering table
\begin{tabular}{ c c c rrrrr} % centered columns (4 columns)
\hline\hline %inserts double horizontal lines
Transition & $\mu/t_1=0$ & $\mu/t_1=0.2$ & $\mu/t_1=1.6$ & Equation:\\ %[0.5ex] % inserts table
%heading
\hline % inserts single horizontal line
-1 $\rightarrow$ 1 & 0 & 2$\mu/t_1$ & 2$\mu/t_1$ & \ref{eq:trans_m1_to_1} \\%[-1ex] % inserting body of the table
-1 $\rightarrow$ 2 & 0 & 0.22 &2.64 & \ref{eq:trans_m1_to_2}\\
-1 $\rightarrow$ 3 & $\sqrt2$ & $\sqrt2$ & $2\mu/t_1-\sqrt2=1.8$ & \ref{eq:trans_m1_to_3}\\%[1ex]

-2 $\rightarrow$ 1 & 0 & 0.77 & 3.8 & \ref{eq:trans_m2_to_1}\\%[-1ex]
-2 $\rightarrow$ 2 & 0 & $2\mu/t_1=0.4$ & 2$\mu/t_1=3.2$ & \ref{eq:trans_m2_to_2}\\ % inserting body of the table
-2 $\rightarrow$ 3 & $\sqrt2$ & $\sqrt2$  & 2.15& \ref{eq:trans_m2_to_3} \\%[1ex]

-3 $\rightarrow$ 1 & $\sqrt2$ & 1.8 & 4.8 & \ref{eq:trans_m3_to_1}\\%[-1ex]
-3 $\rightarrow$ 2 & $\sqrt2$ & 1.64 & 4.05 & \ref{eq:trans_m3_to_2}\\
-3 $\rightarrow$ 3  & 2$\sqrt{2}$ & 2$\sqrt{2}$ & 2$\mu/t_1=3.2$ & \ref{eq:trans_m3_to_3}\\%[1ex]

1 $\rightarrow$ 2  & Not Possible & 0.17  & 0.56& \ref{eq:trans_1_to_2}\\%[-1ex]
1 $\rightarrow$ 3  & Not Possible & $\sqrt2$  & $\sqrt2$ & \ref{eq:trans_1_to_3}\\
2 $\rightarrow$ 3 & Not Possible & $\sqrt2$ & 0.86 & \ref{eq:trans_2_to_3}\\ %[1ex] % [1ex] adds vertical space
\hline %inserts single line
\end{tabular}
\label{table:nonlin} % is used to refer this table in the text
\end{table}

The numerical values of the threshold energies (in units of $t_1$) are given in Table \ref{table:nonlin} where the numerical values correspond to the exact band structure (see Appendix B) because the parabolic / linear approximation becomes to fail for higher values of $\mu$.
It becomes clear from Table \ref{table:nonlin} that almost only the $1 \rightarrow 3$ and $2 \rightarrow 3$ transitions (not possible for $\mu=0$) are responsible for the onset of the resonant peak at $\hbar\omega=\Delta$. We calculated these contributions to the optical conductivity and found that the intensity of the $1 \rightarrow 3$ transition exceeds by far that of the other one. It can be understood by the fact that the dispersion curves $E_1(K)$ and $E_3(K)$ are nearly parallel for a broad range of $K$ values (band nesting) and therefore the joint density of states for this transition is large, as can be seen clearly from Fig.~\ref{fig:f2}.
This figure  takes into account only the $1 \rightarrow 3$ transition and shows that it is responsible for the intensity of the resonant absorption band. The increasing intensity of the $1 \rightarrow 3$ transition as $\mu$ grows is related to the growing number of occupied states that are depopulated by absorbing electromagnetic radiation. This is the principal spectral feature observed in Ref.~\cite {Ubrig2012} (where it was denoted "band C"), which can be considered as characteristic of ABA graphene~\cite{Henriksen2012,Raman} and we return to it below. For heavily doped $p-$type layers studied in Ref.~\cite {Ubrig2012}, two weak absorption bands (denoted there as A and B bands) were observed at lower frequencies and attributed to the $-3 \rightarrow -2$ and $-2 \rightarrow -1$ transitions. In our consideration of $n-$type layers these transitions correspond to $1 \rightarrow 2$ and $2 \rightarrow 3$, respectively, both involving the Dirac-type band. Indeed, they are allowed (even though only for doped samples) but, according to our results, their intensity is rather low (for instance, we cannot see any feature at $\omega /t_1\approx 0.86$, the threshold frequency of $2 \rightarrow 3$ transition for $\mu /t_1=1.6$ (see Table \ref{table:nonlin}). Possibly some further effects can enhance transitions to and from the Dirac-type electronic bands.

The approximately triangular shaped feature in the spectrum  for $\mu=0$ ($\hbar\omega\geq\Delta$) (that could be anticipated institutively~\cite{Orlita-Potemski}) splits into two sub-bands corresponding to the $-1 \rightarrow 3$ and $-3 \rightarrow 1$ transitions. At $\mu=0$, they have the same energy $(=\Delta)$ but for $\mu>0$ the former is shifted to higher energy (see Table 1). Note that, by chance, in Figs.~\ref{fig:f3} and ~\ref{fig:f4} the features related to the $-3 \rightarrow 1$ transition for $\mu /t_1=0.2$ and to the $-1 \rightarrow 3$ one for $\mu /t_1=1.6$ appear nearly at the same frequency, $\omega /t_1\approx 1.8$. Such a feature was not observed in Ref.~\cite {Ubrig2012} (whose experimental situation qualitatively corresponds to our case of $\mu /t_1=0.2$), possibly because it was hindered by Fabri-Perot interference in the substrate.

The intensity of the 1$\rightarrow$3 band as function of $\mu$ can be evaluated analytically if we assume that the matrix element does not depend on $K$:
\begin{eqnarray}
I(\omega)=\frac{const}{\omega^2S}\sum_{\vec{k}}\delta\left\{E_3(\vec k)-E_1(\vec k)-\hbar\omega\right\}\left\{n_f \left[E_1(\vec k)\right]-n_f \left[E_3(\vec k)\right]\right\}\:.
\nonumber
\end{eqnarray}
Changing from sum to integration and using the parabolic approximation (\ref{eq:parab}) near the Dirac point, in the limit of low energies for $E_1(\vec k)$ and $E_3(\vec k)$ we obtain:
\begin{eqnarray}
\nonumber
&& I(\omega) =const\cdot \frac{ m_1}{2\pi(\hbar\omega)^2}{\mathcal L} (\hbar \omega-\Delta)\\
&& \times \int_0^\mu dE_1\left\{\theta\left[\mu-E_1(\vec k)\right]- \theta\left[\mu-E_1(\vec k)- \hbar\omega\right]\right\}\:.
\label{intensity}
\end{eqnarray}
Here ${\mathcal L} (\hbar \omega-\Delta)$ denotes a Lorentzian replacing the $\delta-$ function broadened because of natural reasons.
The integral in Eq. (\ref {intensity}) depends on the value of $\mu$, that is, $I(\omega)\sim \mu$ for $\mu <\Delta$ and $I(\omega)=\mbox
{const}$ for $\mu >\Delta$. This analytical result is compared to the numerical data in Fig. \ref{fig:f5}.

{Higher order hopping parameters, in particular, those connecting more distant atoms in adjacent layers and usually denoted $\gamma _3$ and $\gamma _4$~\cite{kuzmenko,Menezes2014}, are known to cause a distortion (trigonal warping) of the low-energy bands and a small electron-hole asymmetry~\cite{Menezes2014}. Therefore they can slightly affect the intensity and shape of the resonant absorption band but its position, to a good approximation, is determined by the $t_1$ hopping~\cite{Menezes2014}. This is confirmed by the results of our {\it ab-initio} DFT calculations for intrinsic graphene, shown in the inset of Fig.~\ref{fig:f3}.}

\section{Conclusion}
In summary, we calculated the spectral dependence of the real and imaginary parts of the optical conductivity of Bernal-stacked trilayer graphene.
Even though the energy bands of this material look like a superposition of those of a monolayer and a bilayer (with a larger gap between the Dirac point and higher parabolic band), the optical spectra are rich and interesting, especially in the case of gated (doped) graphene because all kinds of interband transitions are allowed. In particular, there is a strong and narrow resonant band cause by transitions between two "nested" parabolic bands. The intensity of this band is controlled by the Fermi level position and attains a maximum for $\mu \approx 1.4 t_1$ (Fig. \ref{fig:f5}), providing a strong light-matter coupling in the atomically thin layer~\cite {Britnell2012}.
{Although the physical origin of the resonant absorption is clear and it could be predicted by just inspecting the band structure of the ABA stacked graphene, its dependence on the Fermi level is not evident without calculations. It would be interesting to show experimentally that the intensity of the resonant absorption attains its maximum for the Fermi energy of $\mu \approx 0.6$~eV. This can be achieved by combining usual doping and application of gate voltage~\cite {NL2014}}. The application of a large gate voltage can slightly modify the band structure but it can be taken into account by adding appropriate (unequal) constant potentials in the Hamiltonian (\ref{Ham}) for atoms belonging to the different monolayers~\cite {Ubrig2012,Avetisyan2009}.
The modulation of the peak absorption can be interesting for optoelectronic devices such as resonant photodetectors with adjustable sensitivity and it is robust with respect to the temperature. We performed calculations also for $T=300$ K and the spectra are very similar to those presented in the figures, except for the obvious broadening of the Fermi steps seen in Fig. \ref{fig:f3}, i.e. the resonant band remains narrow.
Compared to bilayer graphene, the resonant band occurs at a (1.4 times) higher frequency, which broadens the range of applications of Bernal-stacked multilayer graphene materials that can be directly exfoliated from natural graphite.
Also, we would like to point out that the Drude conductivity is three times higher than for monolayer graphene with same doping level, accordingly, the surface plasmon frequency would be also higher (by a factor of $\sqrt 3$)~\cite {PRIMER}, extending the spectral range of possible applications of graphene plasmonics.

\section*{Acknowledgements}

Financial support from the COMPETE Programme (FEDER) and the Portuguese
Foundation for Science and Technology (FCT) through Projects
PEst-C/FIS/UI0607/2013 and PTDC/FIS/113199/2009 is gratefully acknowledged.
ZR thanks the hospitality of the Physics Center of Minho University during her stay in Portugal.
YVB, NMRP, and  MIV acknowledge  support by the EC under Graphene Flagship (contract
no. CNECT-ICT-604391).

%\newpage
\appendix
\section*{Appendix A}
\label{Ap:A}

After transforming the Hamiltonian (\ref {Ham}) to the momentum space [by representing $\vert A_{m},\vec R_{n}\rangle=\frac{1}{\sqrt{N}}\sum_{k}\psi_{A_{m},k}\exp\left(-i\vec{k}\cdot\vec R_{n}\right)$, $\vert B_{m},\vec R_{n}\rangle=\frac{1}{\sqrt{N}}\sum_{k}\psi_{B_{m},k}\exp\left(-i\vec{k}\cdot\vec R_{n}\right)$, $m=1,2,3$ and $N$ is the number of unit cells], its matrix form is:
\begin{eqnarray}
\hat{H}=
\left(
\begin{array}{cccccc}
  0 & \phi& 0 & 0 & 0 & 0  \\
  \phi^{*}& 0 & t_1 & 0 & 0 & 0  \\
  0 & t_1 &0 & \phi  & 0 & t_1 \\
  0&  0& \phi^{*}& 0 & 0 & 0 \\
  0&  0& 0& 0 & 0 & \phi \\
  0&  0& t_1& 0 & \phi^{*} & 0
\end{array}
\right)\:,\label{TB}
\end{eqnarray}
where
\begin{equation}
\phi(\vec{k})=-t_{0}\sum_{\delta_{j}}\exp(i\vec{k}\cdot \vec{\delta_{j}})\:,
\nonumber
\end{equation}
which is given explicitly by Eq. (\ref{eq:phi}). Note that the relation of this $\phi$ to the frequently defined~\cite{neto,Peres-RMP}
quantity $f(\vec{k})$ is the following: $\vert \phi(\vec{k})\vert^2=3+f(\vec{k})$. The Hamiltonian (\ref{TB}) is written in the basis of the atomic orbital eigenfunctions:
\begin{eqnarray}
\psi_{\vec k}=
\left(
\begin{array}{cccccc}
  \psi_{A_1,k} &\psi_{B_1,k} & \psi_{A_2,k} & \psi_{B_2,k} & \psi_{A_3,k} &  \psi_{B_3,k}\\
\end{array}
\right)^{T}\:.
\end{eqnarray}

The ABA trilayer is mirror symmetric with respect to the middle layer,
therefore one can transform the Hamiltonian into a tridiagonal form using a
unitary transformation~\cite{duppen,koshino1}
\begin{eqnarray}
\hat{U}=
\left(
\begin{array}{cccccc}
  \frac{1}{\sqrt2} & 0 & 0 & 0 & -\frac{1}{\sqrt2} & 0  \\
   0 & \frac{1}{\sqrt2} & 0 & 0 & 0 & -\frac{1}{\sqrt2}  \\
  0 & 0 &1 & 0  & 0 & 0 \\
  0&  0& 0 & 1 & 0 & 0 \\
  \frac{1}{\sqrt2} & 0 & 0 & 0 & \frac{1}{\sqrt2} & 0  \\
   0 & \frac{1}{\sqrt2} & 0 & 0 & 0 & \frac{1}{\sqrt2}
\end{array}
\right)\:.\label{eq:transf-Hamiltonian}
\end{eqnarray}
In the new basis the Hamiltonian reads
\begin{eqnarray}
\hat{H}^\prime=\hat{U}\hat{H}\hat{U}^{-1}=
\left(
\begin{array}{cccccc}
  0 & \phi& 0 & 0 & 0 & 0  \\
  \phi^{*}& 0 & 0 & 0 & 0 & 0  \\
  0 & 0 &0 & \phi  & 0 & \sqrt2t_1 \\
  0&  0& \phi^{*}& 0 & 0 & 0 \\
  0&  0& 0& 0& 0 & \phi \\
  0&  0& \sqrt2t_1& 0 & \phi^{*} & 0
\end{array}
\right)\:.\label{TB-Hamiltonian}
\end{eqnarray}
This matrix is composed of two blocks containing only intralayer hoppings and an effective interlayer one, ${\sqrt2 t_1}$.
It can easily be diagonalised yielding Eqs. (\ref{eq:e1} - \ref{eq:e3}).

\section*{Appendix B}
\label{Ap:B}
Here we derive the expressions for the characteristic energies presented in Table 1, beyond the parabolic (linear) approximation for the band spectra.
First we consider the case where $\mu=0.2t_1$, i.e. lies below the bottom of the band $E_{+3}(\vec{k})$.
Among the first three possible transitions from the $E_{-1}(\vec{k})$ to the upper bands, the ${-1 \rightarrow 3}$ transition threshold remains the same [see Eq. (\ref{eq:trans_m1_to_3})] and similar to the case of $\mu=0$.
Transition from ${-1 \rightarrow 1}$ become possible if $\hbar\omega>2\mu$, then for $\mu=0.2t_1$ the allowed energies are $\hbar\omega> 0.4t_{1}$.

The domain of allowed energies for the the transition ${-1 \rightarrow 2}$
can be obtained as $\hbar\omega>E_{+2}-E_{-1}=\sqrt{\vert \phi\vert^2}+\sqrt{t_{1}^2+\vert \phi\vert^2-t_{1}\sqrt{t_{1}^2+2\vert \phi\vert^2}}$. Since $E_{+2}=\sqrt{\vert \phi\vert^2}=\mu$, we obtain
\begin{eqnarray}
\hbar\omega>\mu+\sqrt{t_{1}^2+\mu^2-t_1\sqrt{t_{1}^2+2\mu^2}}=0.22t_{1},
\label{eq:2}
\end{eqnarray}
the number corresponds to $\mu=0.2t_1$.

For the second group of transitions from the $-2$ band, the $-2 \rightarrow 3$ threshold remains as given by Eq. (\ref{eq:trans_m2_to_3}).
The next one, ${-2 \rightarrow  2}$, can occur for energies higher than $2\mu$, namely $\hbar\omega> 0.4t_{1}$. Transition ${-2\rightarrow  1}$ is allowed if $\hbar\omega>E_{+1}-E_{-2}=\sqrt{t_{1}^2+\vert \phi\vert^2-t_{1}\sqrt{t_{1}^2+2\vert \phi\vert^2}}+\sqrt{\vert \phi\vert^2}$. Since $\mu=E_{+1}=\sqrt{t_{1}^2+\vert \phi\vert^2-t_{1}\sqrt{t_{1}^2+2\vert \phi\vert^2}}$ , by solving this equation for $\vert \phi\vert^2=\mu^2+\sqrt2t_1\mu$ we obtain the relation:
\begin{eqnarray}
\hbar\omega>\mu+\sqrt{\mu^2+\sqrt2t_{1}\mu}=0.77t_1.
\label{eq:1}
\end{eqnarray}

Finally, for the last group of three possible transitions from $E_{-3}$ to the three upper bands, for the ${-3 \rightarrow 3}$ one Eq. (\ref{eq:trans_m3_to_3})] holds beyond the parabolic approximation, ${-3 \rightarrow 1}$ is allowed if $\hbar\omega > \sqrt2t_1+2{\mu}$ (then $\hbar\omega> 1.8t_{1}$), and the ${-3\rightarrow 2}$ transition is allowed if
\begin{eqnarray}
\hbar\omega>\mu+\sqrt{t_1^2+\mu^2+t_1\sqrt{t_1^2+2\mu^2}}=1.64t_1.
\label{eq:9}
\end{eqnarray}

The non-zero chemical potential gives rise to new transitions that are not possible in the intrinsic material. The ${1 \rightarrow  3}$ transition is allowed {for the frequencies in the vicinity of} $\omega\equiv \sqrt2 t_{1}/\hbar$.
 The ${1 \rightarrow  2}$ one occurs in the domain
\begin{eqnarray}
\mu-\sqrt{t_{1}^2+\mu^2-t_1\sqrt{t_{1}^2+2\mu^2}}<\hbar\omega<\sqrt{\mu^2+\sqrt{2}\mu t_1}-\mu,
\label{eq:2a}
\end{eqnarray}
For $\mu=0.2t_1$ this corresponds to $0.17t_{1}<\hbar\omega<0.37t_{1}.$

The situation is similar for the ${2 \rightarrow 3}$ transition because, in addition to a threshold energy, there is also a cut-off:
\begin{eqnarray}
\sqrt{\mu^2+t_1^2+t_1\sqrt{t_1^2+2\mu^2}}-\mu<\hbar\omega<\sqrt2 t_1\!,
\label{eq:2aa}
\end{eqnarray}
where the expression in the left hand side is equal to $1.24t_1$ for $\mu=0.2t_1$.

The case of large chemical potential, ($\mu=1.6t_1$), above the bottom of the band $E_{+3}$ corresponds to the situation where the third band is partially filled. For $\hbar\omega>2\mu$, the $-1 \rightarrow 1$ transition becomes possible. The frequency domain for transition $-1 \rightarrow 2$ is similar to the previous case [Eq. (\ref{eq:2})] but for $\mu=1.6t_1$ we have $\hbar\omega>2.64t_{1}$.

Transition from $-1$ to $3$ is allowed for $\hbar\omega>E_{+3}-E_{-1}=\sqrt{t_{1}^2+\vert \phi\vert^2+t_{1}\sqrt{t_{1}^2+2\vert \phi\vert^2}}+\sqrt{t_{1}^2+\vert \phi\vert^2-t_{1}\sqrt{t_{1}^2+2\vert \phi\vert^2}}$. From the dispersion relation $\mu=E_{+3}$ we obtain $\vert \phi\vert^2=\mu^2-\sqrt2t_1\mu$, and $\hbar\omega>2\mu-\sqrt2t_{1}$.
Notice that this equation is exactly the same as Eq. (\ref{eq:trans_m1_to_3}), obtained within the parabolic approximation.

Among the next three possible transitions, the energy domain for $-2 \rightarrow 1$ is given by Eq. (\ref{eq:1}). For $\mu=1.6t_1$ we have $\hbar\omega> 3.8{t_{1}}$. The $-2 \rightarrow  2$ transition is possible when $\hbar\omega> 2\mu$.
For  ${-2 \rightarrow  3}$, from the dispersion relation $\mu=E_{+3}$ we have $\vert \phi\vert^2=\mu^2-\sqrt2t_1\mu$. Hence, $\hbar\omega>E_{+3}-E_{-2}=\mu+\sqrt{\mu^2-\sqrt{2}t_1\mu}=2.15t_1$.

The transitions $-3 \rightarrow 1$ and $-3\rightarrow 3$ are allowed if $\hbar\omega>2\mu+\sqrt2t_1=4.8{t_{1}}$  (similar to previous case) and $\hbar\omega> 2\mu$, respectively.
Transition from $-3$ to $2$ is described by Eq. (\ref{eq:9}), then $\hbar\omega> 4.05t_{1}$.

Finally, $1\rightarrow2$ is described by Eq. (\ref{eq:2a}) (then {$0.56t_1\le\hbar\omega\le 0.6t_1$}); $1 \rightarrow   3$ becomes possible for $\hbar\omega\equiv \sqrt2 t_1$ and $2\rightarrow3$ for the finite frequency domain defined by  $\sqrt{t_{1}^2+\mu^2+t_1\sqrt{t_{1}^2+2\mu^2}}-\mu<\hbar\omega< \mu-\sqrt{\mu^2-\sqrt2t_{1}{\mu}}$. For $\mu=1.6t_1$ it means $0.86t_{1}<\hbar\omega<1.06t_{1}.$

%---------------------------------------------------------------------------------------
%	REFERENCES
%---------------------------------------------------------------------------------------

\end{document}